\documentstyle[11pt,paspconf]{article}
\input epsf
\newcommand{\problem}[2]{ \noindent {\sl Problem }[#1] {\sl #2} \par }
\begin{document}
\title{The Age of the Universe}
\author{Raul Jimenez} 
\affil{Royal Observatory, Blackford Hill, Edinburgh EH9 3HJ, UK}
\begin{abstract}
In this article I review the main methods for determining the age of the 
Universe. I describe how to determine the age of the oldest known 
systems at $z=0$, the system of galactic globular clusters, using 
different techniques. I also describe how to date the Universe 
using the decay of radioactive elements (Cosmochronology). Finally, I 
focus on how to determine the age of the Universe at different 
redshifts and specially the age of radio-quiet galaxies at high redshift. I 
finish by arguing that the {\it most probable age for the Universe is 
$14 \pm 2$ Gyr}.
\end{abstract}
\keywords{globular clusters, cosmochronology, age of the universe}

\section{Introduction}
One of the most important questions in Cosmology is to know the age of 
the Universe. Unfortunately, a definitive answer is still to be found.

Once the values of $\Omega_0$ and $H_{0}$ have been established for 
our preferred cosmological model, the age of the Universe is unequivocally  
determined. In the case of an Einstein-de Sitter Universe  
$t=(2/3)H_{0}^{-1}$ (see e.g. Padmanabham 1993 for cases where 
$\Omega \ne 1$ and $\Lambda \ne 0$). On the contrary, if we are able 
to estimate the age of the Universe by other methods (e.g. dating 
stellar systems) then we can constrain the cosmological model and 
determine which kind of Universe we are living in. We can repeat 
the above argument at different redshifts, in this way we will gain 
information not only about $H_0$ but also about $\Omega_0$ since, e.g., at 
$z=1.5$ the age of the Universe is $t=1.6h^{-1}$ Gyr if $\Omega_0=1$ and 
$t=2.6h^{-1}$ Gyr if $\Omega_0=0.2$, where $h=H/100\, \rm km s^{-1} Mpc^{-1}$. 
Therefore, an {\it accurate} knowledge 
of the age of the Universe at different redshifts can tell us what 
kind of Universe we live in.

Stars are the only clocks in the Universe available to us that are 
``cosmology'' independent. The evolution of stars depends only on the 
rate at which the different nuclear burnings take place. So a 
good strategy to measure the age of the Universe is to find the oldest 
stars at every redshift. 

This article is organised as follows: In section \ref{raul:stellar}, 
I review the stellar 
evolution of low mass stars. A detailed 
review on how to estimate globular clusters (GC) ages follows in the next 
section. Cosmochronology is 
the subject of the next section. In section \ref{raul:galaxies},  
I describe how to date 
high-redshift galaxies. I finish with a summary of the different ways 
to date the Universe.

\section{Stellar evolution of low-mass stars} \label{raul:stellar}
This  section is devoted to reviewing the principles of stellar evolution for 
stars whose mass is below about 2.0 $M_{\odot}$ and above 0.5 $M_{\odot}$. 
These stars are the ones of interest when dating the oldest objects 
at $z < 4$ since 
their lifetimes are between 1 and 20 Gyr. The evolution of a low mass 
star has the following stages.

\begin{itemize}

\item Fragments of the interstellar medium collapse until they reach 
hydrostatic equilibrium.

\item The star now suffers a quasi-static gravitational contraction until 
the free fall is stopped by hydrogen burning at the center. The so-called 
{\it main sequence} starts and H is transformed into He. This is the longest 
period in the life of the stars and it lasts until H is exhausted in the core.

\item After H has been exhausted at the centre of the star, it continues 
to be burned in a very thin shell around the ashes of the previous H 
burning core. The He core contracts and becomes degenerate and the envelope 
expands. The star goes into the red giant phase.

\item As the star moves up in luminosity and the surface cools, mass is lost 
 from the uppermost layers. The degeneracy at the core is 
removed suddenly when He to C/O burning starts. This is the helium flash and 
the red giant branch (RGB) evolution is finished at this point.

\item After the helium core flash has taken place the star goes into the 
horizontal branch (HB) where it will burn He into C/O until He is 
exhausted in the He-core.

\item The star now initiates 
the second ascent to the giant branch, this time 
with a double shell burning around a degenerate core of C/O. These two shells
 :  H $\rightarrow$ He, He $\rightarrow$ C/O, continue to burn until rapid 
detachment of a considerable fraction of the remaining envelope is produced 
 (this is the Planetary Nebula phase). After this the core contracts until 
it becomes degenerate yet again and cannot contract anymore (the star 
becomes a white dwarf with a degenerate C/O core) and the star will follow 
the white dwarf cooling sequence.

\end{itemize}

\begin{figure}
\centering
\leavevmode
\epsfxsize=10cm
\epsfbox{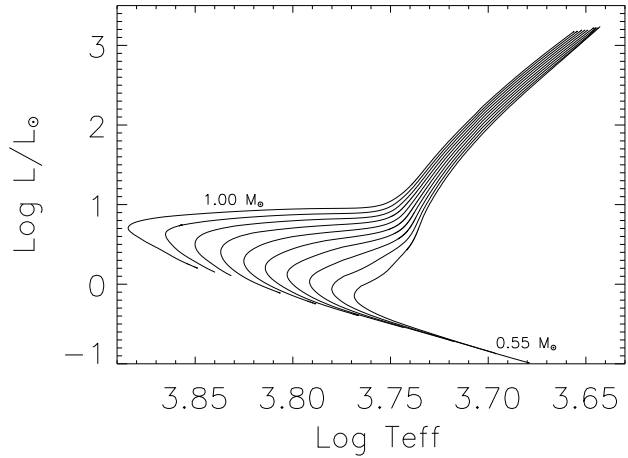}
\caption[]{The effect of different choices of the total mass on the
evolutionary tracks is shown for metallicity  $Z=0.0002$, helium content 
$Y=0.24$ and mixing length $\alpha=1.5$. The
turnoff is markedly affected and its position depends on the mass of the 
star, and therefore on its age.} \label{raul:raulfig2}
\end{figure}

This sequence is well represented when the evolution of a star is plotted 
in the plane $T_{\rm eff}$ {\it vs} Luminosity ($L$). 
In Fig.~\ref{raul:raulfig2} 
this evolution 
is shown for a set of stellar masses in the range 1.0 to 0.55 $M_{\odot}$. 
The most remarkable fact is that the location of the main sequence 
turn off (MSTO) changes with 
mass (i.e. with {\it age}). 
Therefore it should be possible in principle to 
measure the age of a {\it coeval} stellar population if its MSTO is 
{\it well defined}. 
In the figures we show the evolution from the zero age main 
sequence (ZAMS) up to the red giant branch tip (RGT). 
 All tracks have been computed using the latest version of JMSTAR 
(Jimenez \& MacDonald 1997).
A very comprehensive and detailed review of stellar evolution is given by, e.g. 
 Kippenhahn \& Weigert (1990) and Hansen \& Kawaler (1994).

The important point to notice here is that stars are a {\it natural} clock 
in the Universe and that they experience {\it noticeable} changes during 
their life-time and therefore provide an {\it independent} method to compute 
the age of the Universe. 

\section{Globular Cluster Ages}
We have to focus our attempts to compute stellar ages 
in {\it suitable} stellar 
populations (ie. suitable clocks) . By {\it suitable} we mean the following:
\begin{itemize}
\item All the stars must be born at the same time.
\item The population must be chemically homogeneous.
\item There must have been no further episodes of star formation 
that gave birth to new stars which could cover up the {\it oldest} 
population.
\end{itemize}
In the Milky Way there is only one set of such systems: 
the Globular Clusters (GC).

\begin{figure}
\centering
\leavevmode
\epsfxsize=10cm
\epsfbox{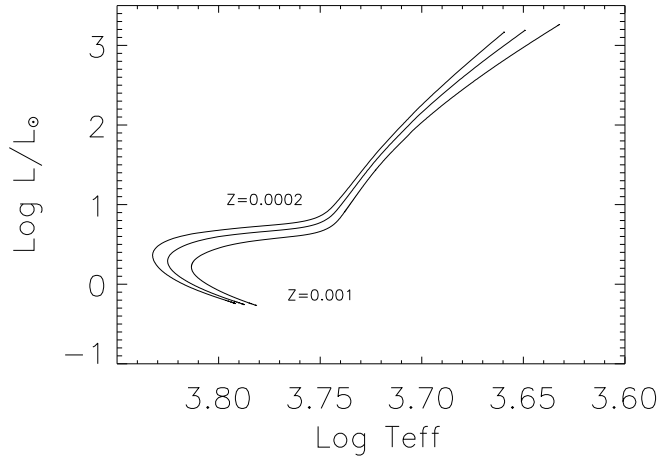}
\caption[]{ Evolutionary tracks are shown for three values of the 
metallicity ($
Z=0.001, 0.0005$ and $0.0002$)
for the same value of the mass (0.8 M$_{\odot}$) and helium content ($Y=0.24$).
The effect of the metallicity is to shift the horizontal position of the 
track.
The higher the metallicity, the higher the opacity and therefore the cooler 
the atmosphere of the star.}
\label{raul:raulfig3}
\end{figure}

GCs are the first places where stars are born in a galaxy (also the bulk 
of the halo stars). There are several mechanisms proposed to explain 
the origin of GCs (Fall \& Rees 1985, Harris \& Pudritz 1994, Padoan, 
Jimenez \& Jones 1997) but all agree that their origin was 
primordial. 
Whatever their origin, the important feature of GCs is that they fulfill 
the above requirements about a {\it suitable} clock. 

\begin{figure}
\centering
\leavevmode
\epsfxsize=8.5cm
\epsffile{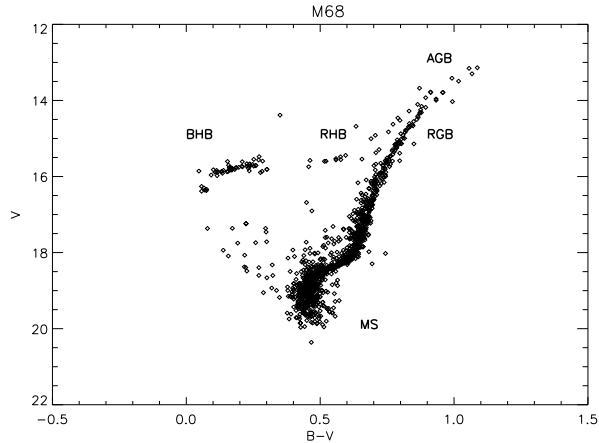}
\caption{ The figure shows the main components in the color--magnitude diagram
for a globular cluster. After stars leave the main sequence (MS) they become
red giants, they follow the path of the red giant branch (RGB). After
degeneracy is removed by the core -- helium core flash -- the stars will
go to the horizontal branch (HB). After this phase the star will start
a fast evolution along the asymptotic giant branch (AGB) where nuclear
burning occurs in a double shell.
} \label{raul:raulfig1}
\end{figure}

There is strong observational evidence that the oldest GCs are chemically 
homogeneous. One of the strongest arguments for this 
is the {\it thin} red giant branches that some GCs have. In 
Fig.~\ref{raul:raulfig3} we 
show the effect of changing the metallicity in a stellar track. It transpires
from the figure that if GCs were chemically inhomogeneous, then the RGB 
should be scattered in the horizontal direction. Fig.~\ref{raul:raulfig1} 
shows the 
colour-magnitude diagram (CMD) 
for M68 (one of the oldest globular clusters). Its RGB is quite 
thin, indicating a homogeneous chemical composition (this is 
further confirmed when spectra of stars are taken in the clusters). 
 The above argument assumes that the mixing length parameter 
is the {\it same} for all stars in the cluster (see below for further 
details).

Support for GCs being old comes from two facts: their metallicity is as 
low as 1/100 of solar and the characteristics of their CMD are those 
corresponding to ages larger than 10 Gyr. 

\subsection{The Isochrone fitting method}
The first (and more obvious method) to compute the age of a GC is to 
exploit the fact that the locus of the MSTO in the plane $T_{\rm eff}$ vs. $L$ 
changes with age (mass). In this 
way one computes different isochrones, i.e. tracks in the plane 
$T_{\rm eff}$ vs. $L$ at the same time for all masses, with the chemical 
composition of the GC and finds the better fit to the MSTO region. To 
do this a very important step is needed: the {\it distance} to the 
GC is needed in order to transform the theoretical luminosity into observed 
magnitudes in different bands, and thus is where trouble starts.
If the distance to the GC is unknown, there is a degeneracy between age and 
 distance. 
In this way, we can simulate a different age by simply getting the GC 
closer or more distant to us.

\problem{1}{Use any set of published isochrones (e.g the Yale isochrones) 
to demonstrate the previous statement. Also give the error that propagates 
in the age determination if the distance modulus is not known better 
than 0.1 mag}

Distances to GCs are very poorly known since it is impossible to get 
the parallax of individual stars and therefore ages of GCs are not 
accurately known using the isochrone fitting method. Usually, there are
different methods to compute distances to GCs: the RR-Lyrae method; the 
subdwarf fitting method; the tip of the red giant branch; and the 
luminosity function. The RR-Lyrae method consists in using the known 
 Period-Luminosity relation for the RR-Lyrae pulsators in the HB, it 
gives an uncertainty of 0.25 mag in the distance modulus determination, 
which translates to 3 Gyr error in the age determination. The subdwarf method 
uses the nearby low metal subdwarfs to calibrate the distance of GCs; again 
its uncertainty is about 0.2 mag. The tip of the RGB method uses the fact 
that stars at the tip of the RGB flash at a well defined luminosity (Jimenez
et al. 1996); therefore the tip of the RGB is well defined and can be 
used as a distance indicator. The luminosity function method is explained 
later on. It gives more precise distance determination, and the error 
in the distance is only 0.05 mag.

In order to circumvent the need for the distance determinations in 
computing the age, Iben \& Renzini (1984) proposed an alternative 
method for deriving ages using the MSTO, the so-called $\Delta V$ method. 
The method exploits the fact 
that the luminosity of the MSTO changes with mass (age) and not only 
its $T_{\rm eff}$ (see Fig.~\ref{raul:raulfig2}), and also that 
the luminosity of the 
 (HB) does not change since the core mass of the 
He nucleus is the same independently of the total mass of the stars 
(provided we are in the low mass range, see section 1), and the luminosity 
in the HB is provided by the He core burning.
Since the method is based on a relative measure (the distance between the 
HB and the MSTO), it is distance independent. Of course, the method 
needs the knowledge of at least one GC distance in order to be zero 
calibrated. Unfortunately, the method has a serious disadvantage: the 
need to know {\it accurately} the location of the MSTO point. This 
turns out to be fatal for the method since it has associated an error 
of 3 Gyr in the age determination.

Furthermore, all the above methods are affected by three main diseases: 
the calibration colour-$T_{\rm eff}$; the calibration of the mixing-length 
parameter; and the need to fit morphological features in the CMD (i.e. 
the MSTO). See Table~\ref{raul:raultable1} for a detailed review of all 
errors involved in the different methods.

The most common ages obtained for the oldest GCs using the MSTO method are 
in the range 14-16 Gyr. Nevertheless, an error bar of 3.5 Gyr is associated 
with all age determinations using the MSTO methods described above.
                                                                               
\subsection{The Horizontal Branch Morphology Method}
A method that is independent of the {\it distance modulus} can be developed
using the fact that the spread of stars along the HB is mainly due to 
previous mass loss which varies stochastically from one star to another. 
The range of colours in which zero age horizontal branch stars (ZAHB) 
are found is a function of metallicity (the `first parameter') and of 
the range of ZAHB masses. More precisely, the ZAHB colour at a given 
metallicity depends on both the total mass of the star and the ratio 
of core mass to total mass, but the core mass is essentially fixed 
by the physics of the He flash and is quite insensitive to the mass 
and metallicity. For a given average mass loss, the average final mass 
is thus a decreasing function of age, which is therefore a popular 
candidate for the `second parameter' (Searle \& Zinn 1978), although other 
candidates such as CNO abundance have also been suggested. A strong case
for age as the chief (though perhaps not necessarily the only) second 
parameter has been made by Lee, Demarque \& Zinn (1994), who find a 
tendency for the clusters to be younger in the outer Galactic halo. 
Jimenez et al. (1996) used analytical fits to a variety of RGB models and 
followed evolution along the RGB with mass loss treated by the Reimers (1975) 
formula. They showed that, for clusters with narrow RGBs (the majority), 
star-to-star variations in initial mass, metallicity, mixing-length parameter
or delayed He core flash at the RGT can be ruled out as a source of the 
spread along the HB. This leaves variations 
in the Reimers efficiency parameter $\eta$ as the only likely alternative. 
Jimenez et al. (1996) 
 demonstrated that, in fact, the origin of the spread of the HB is 
 star-to-star variations in $\eta$ (or some equivalent parameter).

It is therefore meaningful to proceed to an analysis of both the RGT 
and the HB and to link them together to deduce general properties from 
morphological arguments.

The procedure that we use to analyse the morphology of the RGB and the HB 
together and constrain the mass of the stars at the RGB is as follows.

\begin{itemize}

\item The mass on the upper part of the RGB is determined from the average 
mass of the HB, the average mass-loss efficiency and its dispersion. 
Calculating the average mass and then the $2\sigma$ value of the 
distribution will give the range of masses along the HB.

\item Since the vertical position of the RGB depends only on metallicity and 
$\alpha$, once the metallicity is known $\alpha$ is the only free parameter. 
Therefore, we can find a fit for the best value of $\alpha$, using the vertical 
position of the RGB.   

\item Now we have all the necessary parameters to model the RGB and the 
HB. With these data we can calculate a track and give the age at the RGT, 
and therefore the age of the GC itself.

\end{itemize}

In Jimenez et al. (1996) we analysed eight GCs using the above method 
and found that the oldest GCs were not older than {\bf 14 Gyr}. 

\subsection{The Luminosity Function Method}
When the MSTO method is used (both isochrone fitting and $\Delta V$), 
 a different distance modulus
can be mimicked with a different mass for the MSTO, 
and therefore with a different 
age. This age-distance degeneracy in general leads to an uncertainty
in the age of $3$ Gyr. On top of this, other uncertainties in the stellar
physics lead to an additional uncertainty in the age of about $2$ Gyr.

In order to tackle this problem, an alternative method that is independent of
the distance modulus has been proposed by Jimenez et al. (1996).
Here I review the  use of the stellar luminosity function (LF) of
GCs, in order to break the age-distance degeneracy.

The LF seems to be the most natural observable to try
to constrain both age (Paczynski 1984, Ratcliff 1987) and distance
modulus at the same time.
The LF is a natural clock because the number of stars in a given luminosity
bin decreases with time, since more massive stars
evolve more rapidly than less massive ones. The fact that small differences
in stellar masses corresponds to large differences in evolutionary time
explains the power of the LF clock, rather than being a source of uncertainty
in getting GC ages (as it is in the MSTO method).

The LF is also a natural distance indicator, because
the number of stars in a given luminosity bin depends on the position
of the bin.

Stars of different mass evolve along the main sequence at different
speed: the more massive the faster. This means that the number of
stars inside a fixed luminosity bin decreases with time.
This effect is particularly strong around the subgiant region,
so that the whole shape of the LF is changing with time, and
not only its normalization.
In other words {\it the ratio between the number of stars in two different
bins in the LF can be used as a clock for GC ages} (Jimenez \& Padoan 1996).

\begin{figure}
\centering
\leavevmode
\epsfxsize=10cm
\epsfbox{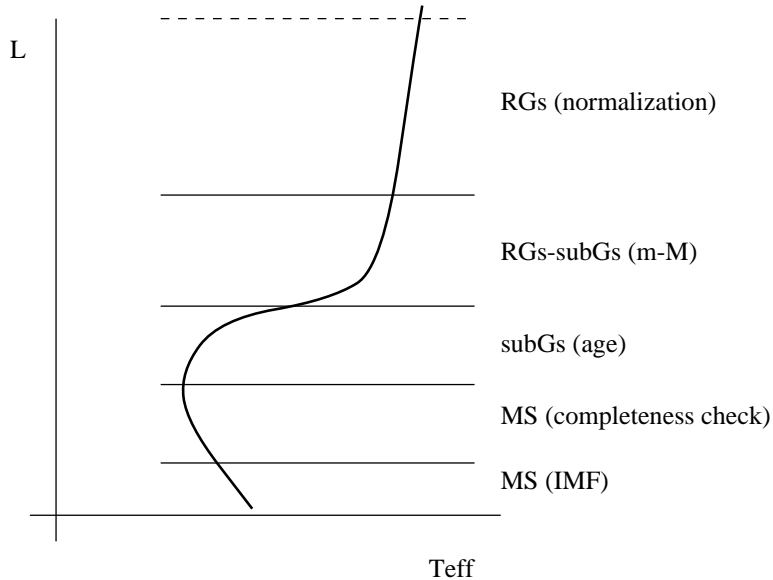}
\caption[]{The figure shows the bins used in the LF method (see text) to 
determine age and distance of a GC}
\label{raul:raulfig7}
\end{figure}

For a determination of both distance and age, one needs to get from
the LF at least two independent constraints, which means
three bins in the LF, since one is required for the normalization. A forth
bin is also very useful in order to check for the completeness of the
stellar counts (see Fig.~\ref{raul:raulfig7}).

Therefore {\it the LF method for determining age and distance of GCs
consists in the production of 4-bin theoretical LFs for GCs, to be
compared with their observational counterparts}.  This comparison is
made for a given chemical composition, that is assumed to
be known from other methods, like spectroscopy.

The number of bins should not be larger than necessary (four), since
each bin should be as wide as possible, in order to reduce the
statistical errors in the stellar counts, due to uncertainty
in the photometry and to the stochastic nature of the stellar mass
function.
The bins we use are all $1$ mag wide, apart from the first one, at the RGB,
that is used for the normalization, and may be extended as luminous as possible
along the RGB.

The second bin, that is the main constraint for the distance modulus,
is positioned between the RGB and the SGB (sub-giant branch),
in order to partially contain the steepest section of the LF
(this gives the sensitivity to a translation in magnitude.).
The third bin, that is the main constraint for the age, contains the SGB,
because this is the part of the LF that is most sensitive to age.
The fourth bin is just next to the third one, and will typically include the
upper part of the main sequence.

The procedure to obtain the LF from evolutionary
stellar tracks is illustrated in Jimenez and Padoan (1996).
A power law stellar mass function is assumed here, as
in that work.

We have shown that a careful binning of the stellar LF
allows a very precise determination of age and distance of GCs, 
at the same time.

If stellar counts with $5\%$ $1\sigma$ uncertainties in $1$ mag wide bins are
available, the age can be determined with an uncertainty of $0.5$ Gyr, and
the distance modulus with an uncertainty of $0.06$ mag.

This LF method is therefore an {\it excellent} clock for relative ages of GCs, 
and
also a very good distance indicator. In other words, its application will
provide very strong constraints for the theory of the formation of the Galaxy.

\begin{figure}
\centering
\leavevmode
\epsfxsize=10cm
\epsfbox{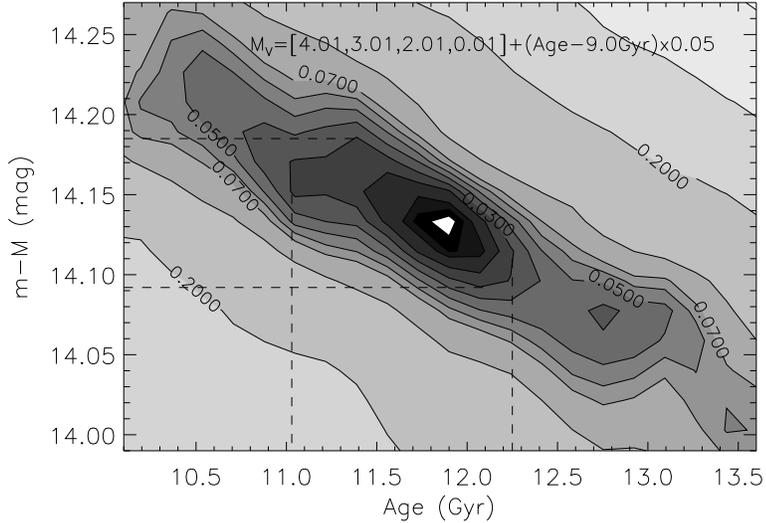}
\caption[]{Distance and age for the low metallicity globular cluster M55 
{\bf simultaneously} determined using the LF method (see main text). The 
contour lines show that the age can be determined with 1 Gyr accuracy and the 
distance modulus with 0.05 provided the number of stars in the LF can be 
complete to a level of 4\%.  
} \label{raul:raulfig5}
\end{figure}

In Fig.~\ref{raul:raulfig5} we show the result of applying the LF method 
to the metal poor galactic globular cluster M55. The plot shows contour plots 
for the error in the determination of the distance modulus and age of M55 
simultaneously. The contour plots correspond to different values for 
the uncertainty in the number of stars in the luminosity function. As stated 
above, if stellar counts are within an uncertainty of 5\% then the age 
is determined with an uncertainty of $0.5$ Gyr, and the distance modulus 
with an uncertainty of $0.06$ mag. 

{\it The age obtained for M55 (12 Gyr) confirms the conclusion of the 
HB morphology method that GCs are not older than 14 Gyr.}

\begin{table}
\begin{center}
\begin{tabular}{cccc}
\hline
 & MSTO & HB & LF \\
\hline\hline
Distance Modulus & 25\% & 0\% & 3\% \\
Mixing Length & 10\% & 5\% & 0\% \\
Colour-$T_{\rm eff}$ & 5\% & 5\% & 0\% \\
He Diffusion & 7\% & 7\% & 7\% \\
$\alpha$-elements & 10\% & 5\% & 10\% \\
Reddening & 5\% & 10\% & 0\% \\
\hline
\end{tabular} 
\end{center}
\caption{The table shows the errors associated with the different methods 
described in the text to compute the age of the oldest globular clusters. The 
first column lists the main uncertainties when computing GCs ages.} 
\label{raul:raultable1}
\end{table}

\section{Cosmochronology}
Cosmochronology uses the radioactive decay of nuclear species to date 
the age of the elements and therefore the life of stellar populations.
The key point is the use of nuclear species that are only produced through 
the r-process, and therefore produced only in massive stars that 
 live for around $10^8$ Gyr. 
The idea is to look at very metal-poor stars and find in their spectrum 
elements that are only produced in the r-process. Then measuring the 
radioactive decay of these elements, it is possible to find the age 
of these very metal-poor stars on the assumption that  
the radioactive elements were 
produced by massive (short-lived) stars in the very early evolution 
of the Galaxy.
In this way the r-process is entirely responsible for the synthesis of 
 $^{187}$Re, $^{232}$Th, $^{235}$U, $^{238}$U and $^{244}$Pu. Knowing 
the short-lived progenitors in alpha decay chains of the above isotopes, one 
can compute the production rates of the pairs $^{232}$Th/$^{238}$U and 
$^{235}$U/$^{238}$U. Using these chronometers it is possible to get the 
age of the Galaxy in the following way:

\begin{itemize}

\item Compute production ratios for the entirely r-process long lived 
isotopes.

\item Get meteoritic abundances ratios of the same isotopes  to obtain the 
duration of nucleosynthesis since the formation of the Galaxy until the 
formation of the solar system.

\item Use a model for the chemical evolution of the Galaxy.

\end{itemize}

Then a variety of predictions are drawn due to the different models used in 
the previous three points. We will simply note here that the latest 
investigations in this subject (Truran, private communication) 
give values between 11 and 14 Gyr for the 
age of the oldest stars in the galaxy. Nevertheless, it should be noted that 
Cosmochronology has an entirely different type of model dependence as compared 
with the uncertainties in GC ages; the results thus are an   
 independent check on the ages of the 
GCs. This therefore adds to the evidence that the most likely age of 
the Universe is around $14 \pm 2$ Gyr.

\section{Ages of High Redshift Galaxies} \label{raul:galaxies}
In the previous sections we have described two independent methods to 
compute the age of the universe at $z=0$. Usually the literature refers 
to the age of the Universe as the age at $z=0$, nevertheless it is possible 
to find at higher redshifts (due to the new 10m class telescopes and 
the {\it Hubble Space Telescope}) {\it suitable} objects where it is  
possible to measure the age of their stars in a {\it reliable} and 
{\it accurate} way. As we have pointed out previously, the age of 
the Universe at different redshifts give us an independent measure 
of $\Omega$. The first point to answer is: are GCs good ``clocks'' at 
high redshift?

Unfortunately not. The reason is obvious; we cannot resolve individual 
stars at high redshift. Therefore the strategy is to find analogues to 
GCs at high redshift bearing in mind that we will have {\it only} 
the integrated light of the high redshift population.

The disks of spirals galaxies are very bad {\it suitable} clocks since we 
know that stars are born continuously (it would make no sense trying to 
ask what is the age of a city by measuring the age of its citizens). 
Much better candidates are elliptical galaxies. Their stars are believe to 
form in a first initial burst, after which there is no further 
significant amount 
gas left to be processed 
into new born stars. Once the stars in a elliptical are born, they 
evolve passively, aging and becoming redder. 
Therefore, ellipticals that have no significant activity from 
active galactic nuclei are the {\bf best} {\it suitable} clocks for 
measuring the age of the Universe at high redshifts.

In this section I describe a method to find ellipticals with no 
significant AGN contribution at high redshift and how to date its population 
from 
their integrated spectrum.    

Finding distant galaxies and analyzing their starlight remains one of the only
direct methods of studying the formation and evolution of galaxies.
Recent studies with the Keck and HST have provided strong evidence
that the majority of star formation took place at relatively recent redshifts
($z \sim 1$) (Lilly et al. 1995).
However this does not mean that
 all galaxies formed at that redshift, and
in fact it seems clear that many galaxies were formed
much earlier at a higher redshift (Driver et al. 1995).

In particular, it is the reddest galaxies at high redshifts which provide the
best constraints on the earliest epochs of galaxy formation and evolution,
since their colour is most likely due to an evolved stellar population.
In a recent publication (Dunlop et al. 1996) we have reported the discovery of
the oldest known galaxy at $z = 1.55$ (53W091).

Once the spectrum of the galaxy has been obtained it is necessary to 
build synthetic stellar populations models in order to compute the 
age of the stellar system.

The necessary steps to compute a synthetic integrated spectrum are the 
following:

\begin{itemize}

\item The first step is to specify the initial mass function and the 
star formation rate of the population that we want to synthesize. Usually 
a Miller-Scalo initial mass function (IMF) (Miller \& Scalo 1979) 
is used (even though the use of a physical IMF 
is more suitable (see Padoan et al. 1997)). The star formation rate 
is different depending on the population. For ellipticals, (and GCs) since 
there is no further episodes of star formation apart from the initial 
burst, an instantaneous burst is a quite good approximation.

\item Select a library of stellar interiors that cover a large range in masses 
(from 0.1 M$_{\odot}$ to 120 M$_{\odot}$).

\item Create an isochrone using the previous IMF and star formation rate (SFR).

\item Using a library of theoretical stellar atmospheres (or observed 
stellar spectra) associate them which every point in the isochrone in order 
to add them all and produce a  synthetic spectrum of the population.

\end{itemize}

A more better and accurate way is to join the last two steps and use 
self consistent stellar evolution models. These models compute  
the interior and the photosphere in one step and therefore for 
every point in the track the proper atmospheric model is known. 
Unfortunately, they are not so easy to compute (see Jimenez et al. 
1997 for a detailed description of these models).

\begin{figure}
\centering
\epsfxsize=13cm
\epsfbox{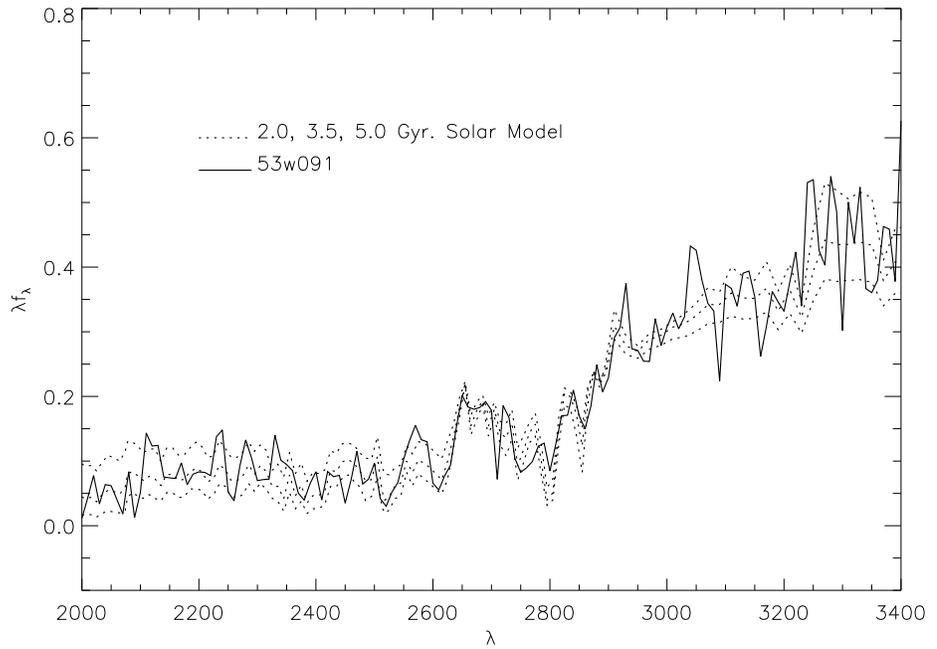}
\caption[]{The spectrum of the high redshift galaxy 53W091 is plotted with 
several synthetic stellar population models for different ages. 53W091 
is the oldest known galaxy at a redshift of $z=1.55$. In order to determine 
its age we have built several synthetic stellar population models at 
different ages. The best fit is found for an age of 3.5 Gyr, the highest 
model corresponds to an age of 2 Gyr and it is to blue to fit the observed 
spectrum of 53W091. 
The lowest model corresponds to an age of 5 Gyr and the stellar population is 
too red (old stars) to fit the observed spectrum.    
} \label{raul:raulfig6}
\end{figure}

In Fig.~\ref{raul:raulfig6} we have plotted the spectrum of 53W091 and three 
different synthetic stellar population models. The effect with age on 
the spectral features is quite well known and the best fit is obtained 
for an age of 3.5 Gyr. 

How robust is this age estimation? We were very careful to chose a ``suitable''
clock at that redshift in order to make a meaningful age determination. 
53W091 is a ``clean'' elliptical (i.e. no contamination from any AGN 
contribution (Dunlop \& Peacock 1993) 
and no new star formation episodes), so the only evolution is 
 due to aging stars. Therefore, it is only a question of having 
accurate stellar models (interior and atmospheric) in order to make accurate 
predictions for the integrated spectrum. Unfortunately, the evolution of 
post-main-sequence stages is not so well understood, and many of the
 differences between different synthetic population models come from their 
different treatment of post-main-sequence stages.

Nevertheless, there are some other features in the spectrum that can help 
us to circumvent these problems and make a robust prediction for the age 
of 53W091.

\problem{2}{Use the breaks around 2600\,\AA\, and 2900\, \AA\, in 
the spectrum of 
53W091 and compute the most probable age of this galaxy. To do so build a 
set of synthetic stellar spectra as described in the text. The main 
advantage of the breaks is that they are reddening independent.}
 
\begin{figure}
\begin{center}
\leavevmode
\epsfxsize=13cm
\epsffile{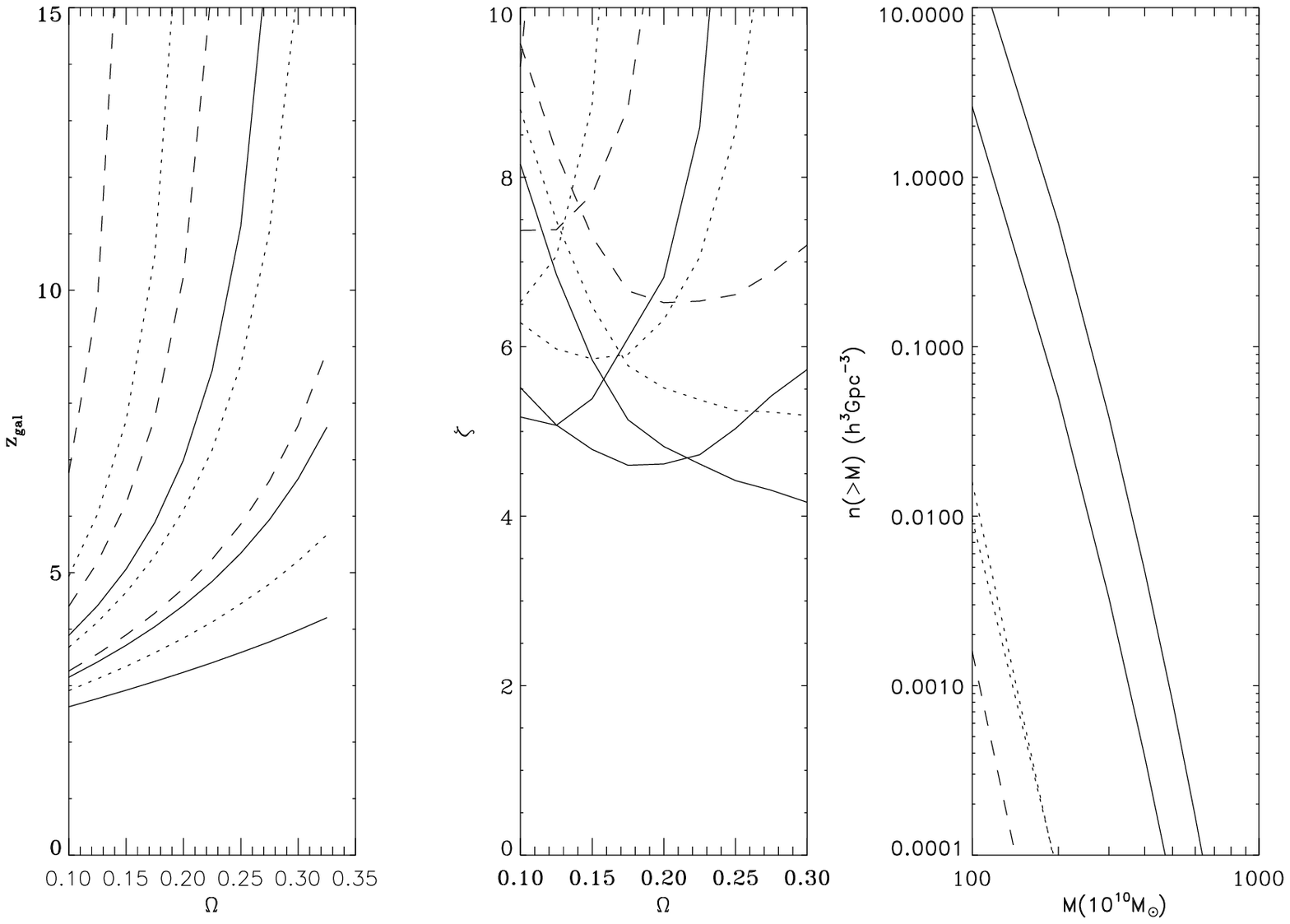}
\caption{In the left panel the redshift, 
$z_{gal}$, when star formation in 53W091 was completed is
plotted vs $\Omega_0$ for $\Omega_0+\Lambda=1$ Universe. Solid lines correspond
to $t_{age}$=3, dotted to 3.5 and dashes to 4Gyr. Three lines of each type
correspond to $h=0.6,\,0.8,\,1$ from bottom up. The middle panel shows $\zeta$, 
the number of standard deviations of the primordial density 
field 53W091 should be in the flat $\Lambda$-dominated CDM models is plotted
vs $\Omega_0$. $\zeta$ scales $\propto b^{-2/3}$ and is plotted for $b$
normalized to the second year COBE/DMR maps.  Same notation as
in previous panel. Finally, in the right panel the predicted number density,
 $n(>M)$, of galaxies like 53W091 with the redshift of formation plotted in 
the left panel is plotted vs their total mass in units of $10^{10} M_{\odot}$.
 Solid lines correspond to $\Omega_0=0.1$:
they are for $h_0=1,\,0.8,\,0.6$ from top to bottom respectively. Dotted lines
correspond to $\Omega_0=0.2$ and $h=0.8,0.6$ from bottom to top; 
the line for $h_0=1$ lies below the box. Dashes correspond to $\Omega_0=0.3$ 
and $h_0=0.6$.} 
\label{raul:raulfig4}
\end{center}
\end{figure}

What are the implications for cosmology of a galaxy aged 3.5 Gyr at $z=1.55$?
We need to know the mass of 53W091 in order to make our cosmological 
implications much stronger. 

In order to compute the mass of an object knowing its apparent magnitude we 
just have to integrate the IMF and scale it until we match the observed 
apparent magnitude. For the case of 53W091 the mass is about $10^{12}$ 
M$_{\odot}$ (see Kashlinsky \& Jimenez 1997). 

We assume in what follows that the data
on 53W091  imply that the galaxy at $z$=1.55 has mass in excess of 
$10^{12}M_\odot$
and its stellar population has an age of $\simeq 3.5$ Gyr. 
What then are the cosmological
implications of at least one object in the Universe having collapsed 
(formed galaxy)
on mass scale of $>10^{12}M_\odot$ at least 3.5 Gyr before 
the redshift of 1.55?

The left box in Fig.~\ref{raul:raulfig4} shows the redshift
$z_{gal}$ at which the galaxy 53W091 must have formed its first stars for
$\Omega$+$\Lambda$=1 Universe.
Solid lines correspond
to $t_{age}=3$Gyr, dotted to 3.5 Gyr and dashed to 4 Gyr. Three types of
each line correspond to
$H_0$=60, 80 and 100 km s$^{-1}$ Mpc$^{-1}$.
One can see that
the value of $z_{gal}$ decreases as {both} $\Omega_0$ and $h_0$ decrease.
On the other hand, in the low-$\Omega_0$ CDM cosmogonies the
small-scale power is
also reduced as the {product} $\Omega h$ decreases;
this would at the same time delay collapse of first galaxies until
progressively smaller $z$.

To quantify this we proceed in the manner outlined in Kashlinsky (1993). This 
involves
the following steps (see Kashlinsky $\&$ Jimenez 1997 for details): 

\begin{itemize}

\item [1)] Specify
the primordial power spectrum, $P(k)$, of the density field at some initial 
redshift
$z_i \gg$1 when the density field is linear on all scales. 
The power spectrum depends
on the initial power spectrum, assumed to be Harrison-Zeldovich, and the 
transfer function
which accounts for the evolution of the shape of the power spectrum in the 
linear regime.
The latter depends in such models only on the product $\Omega h$ and
was adopted from Bardeen et al. (1986).

\item [2)] Compute the amplitude,
$\Delta_8$, of
that field at $z_i$ on the scale of $8h^{-1}$Mpc that produces the observed 
unity rms fluctuation in galaxy counts today, or a $1/b$ amplitude 
in mass fluctuation ($b$ is the bias factor) at $z$=0. 

\item [3)] Compute the fractional density contrast,
 $\delta_{col}(z)$, the fluctuation had to have at $z_i$ in order to collapse 
at $z$. 

\item [4)] A convenient quantity to describe 2) and 3) is $Q(z)\equiv 
\delta_{col}(z)/\Delta_8$. For $\Lambda$-dominated flat Universe and
 in the limit of $1+z_{gal} > \Omega^{-1/3}$ it can be approximated as 
$Q(z)\simeq 3\Omega^{0.225} b^{2/3}(1+z)$. 

\item [5)] $b$ is determined by 
normalizing the density distribution given by $P(k)$ to the COBE-DMR maps 
(Bennett et al 1994; Stompor et al 1995). 

\item [6)] Given $P(k)$ we compute the rms 
fluctuation, $\Delta(M)$, over a region containing
mass $M$. 

\item [7)] The quantity $\zeta \equiv Q(z_{\rm gal}) \Delta_8/ \Delta(M)$ 
then describes the number of standard deviations an object of mass $M$ had to 
be in order to collapse at $z_{\rm gal}$
in the cosmological model specified by $P(k)$.

\end{itemize}

The values of $\zeta$ for $M$=$10^{12} M_\odot$ are plotted versus $\Omega_0$ in
the middle box of Fig.~\ref{raul:raulfig4} 
for $\Omega_0$+$\Lambda$=1 for various
values of $t_{\rm age}$ and $h$. As in the left box solid lines correspond to
$t_{\rm age}$=3 Gyr, dotted to 3.5 Gyr and dashes to 4 Gyr. Three types of each line
correspond to $h$=0.6, 0.8 and 1 going from bottom to top at large values
of $\Omega$ at the right end of the graph. The line for $t_{age}$=4 Gyr and
$h$=1 lies above the box. As the figure shows, this galaxy must
represent an extremely rare fluctuation in the density field specified by the
low-$\Omega$ flat CDM models.
Note that the total mass of 53W091 must be at
least a factor of 10 larger in which case our conclusion will be much stronger.

The right box shows the expected co-moving number density of such galaxies, 
$n(>M)$, in units of ($h^{-1}$Gpc)$^{-3}$ vs the total (dark+luminous) mass for
$t_{age}$=3.5 Gyr. It was computed using the Press-Schechter  
prescription.
The numbers for the co-moving number density
$n(>M)$ were computed for $\Omega_0$=0.1, 0.2, 0.3 and
$h$=1, 0.8, 0.6. Solid lines correspond to $\Omega_0$=0.1 and to
$h$=1, 0.8
from top to bottom. Dotted lines correspond to $\Omega_0=0.2$ and $h_0$=
1, 0.8, 0.6
from top to bottom respectively. The dashed
line corresponds to $\Omega_0=0.3$ and $h_0$=0.6.
The models not shown lie below
the box.
One can conclude from Fig.~\ref{raul:raulfig4} that within the framework of the $\Lambda$ cold dark matter models this object must be extremely rare in the 
Universe. There exists a narrow range of parameters (total mass of $10^{12}
 M_{\odot}$, age less than 3 Gyr, $\Omega=0.1$ and $h \geq 0.8$) for which 
one expects to find a few such objects with each horizon, but for most cases 
the number density of such objects is less than one per horizon volume.

\section{Summary}

In this article we have focused our attention on how to calculate the age 
of the Universe at different redshifts and its cosmological implications.
The main conclusions are the following:

\begin{itemize}

\item There are well defined stellar clocks in the Universe at different 
redshifts that allow us to obtain independent measures of the age of 
the Universe.

\item Galactic Globular Clusters are the best cosmological clocks at 
$z=0$. Several methods have been used to compute their ages. The 
{\it Luminosity Function} is the most accurate method to compute 
the age and distance of Galactic Globular Clusters. {\it The  age 
of the oldest globular clusters is $14 \pm 2$ Gyr}.

\item Cosmochronology gives a most probable age for the oldest 
stars of the Galaxy between 11 and 14 Gyr. 

\item At high redshift quiet radio-galaxies are the best cosmological 
clocks to probe the early evolution of the Universe. I have described the case 
of 53W091 and give detailed description on how to compute its age. This 
galaxy (the oldest at $z=1.55$) was found to be 3.5 Gyr old. It 
represents a very rare event in modified cold dark matter models. 

\end{itemize}

Based on the previous arguments, we can say that the Age of the 
Universe is between {\it 11 and 16 Gyr, with a  most probable age of 14 Gyr}.
This constrains 
the allowed values for $\Omega$. If $H_0=65$ km s$^{-1}$ Mpc$^{-1}$ (see 
this volume) then 
$\Omega \leq 0.2$, otherwise $\Lambda \ne 0$. In the case $\Omega=1$ then 
$H_0 \approx 0.5$ km s$^{-1}$ Mpc$^{-1}$.

\end{document}